\documentclass[]{emulateapj}

\usepackage{graphicx}
\usepackage{wrapfig}
\usepackage{natbib}

\newcommand{\be}{\begin{equation}}
\newcommand{\ee}{\end{equation}}
\newcommand{\nn}{\mbox{} \nonumber \\ \mbox{} }
\newcommand{\ba}{\begin{eqnarray}}
\newcommand{\ea}{\end{eqnarray}}
\newcommand{\om}{\omega}
\newcommand{\Alfven}{Alfv\'{e}n }

\newcommand\eg{{\it{e.g.\ }}}

\newcommand{\Bf}{{magnetic field}}
\newcommand{\Bfs}{{magnetic fields}}
\newcommand{\Ef}{{electric  field}}

\newcommand{\NS}{neutron star}
\newcommand{\NSs}{{neutron stars}}
\newcommand{\EM}{electromagnetic}

\newcommand{\ms}{magnetosphere}
\newcommand{\mss}{magnetospheres}

\newcommand{\LC}{light cylinder}
\newcommand{\Lf}{Lorentz factor}

\begin{document}

\title{Escape of Fast Radio Bursts from magnetars' magnetospheres}
\author{Maxim Lyutikov}
\affiliation{ Department of Physics and Astronomy, Purdue University, 
 525 Northwestern Avenue,
West Lafayette, IN
47907-2036, USA; lyutikov@purdue.edu}

\begin{abstract}
We discuss  dissipative   processes occurring during production and  escape of    Fast Radio Bursts  (FRBs) from magnetars' \mss, the presumed {\it loci} of FRBs.  
High \Bf\ is   required  in the emission region,  both to account for the overall  energetics  of FRBs, and  in order to suppress ``normal'' (non-coherent)  radiative losses of radio  emitting particles; this limits the  emission radii to $\leq {\rm few} \times  10 R_{NS}$. 
Radiative losses by particles in the strong FRB pulse  may occur  in the outer regions of the \ms\ for longer rotation  periods, $P\geq 1$ second.   These  losses
   are suppressed by several effects: (i)  the ponderomotive     pre-acceleration of background plasma  along  the direction of wave propagation (losses reduced approximately as  $\gamma_\parallel^{3}$:  smaller frequency, $ \propto  \gamma_\parallel^2$ in power,   and times scales stretched, $ \propto \gamma_\parallel$); this acceleration is non-dissipative and is  reversed on the  declining part of the pulse; (ii) Landau-Pomeranchuk-Migdal effects (long radiation formation length and ensuing destructive interference of scattered waves).   In some cases an FRB pulse  may be  dissipated on external perturbations (\eg\ an incoming pulse of \Alfven waves): this may   produce a pulse of UV/soft X-rays   possibly detectable by Chandra.
   \end{abstract}


\section{Introduction}

Observations of  correlated radio and X-ray bursts \citep{2020Natur.587...54C,2021NatAs...5..372R,2020Natur.587...59B,2020ApJ...898L..29M,2021NatAs.tmp...54L} 
established the  FRB-magnetar connection. Temporal coincidence between the radio and X-ray profiles, down to milliseconds, argues in favor of magnetospheric origin 
\citep{2020arXiv200505093L}: we know that X-ray are magnetospheric events, as demonstrated by the periodic oscillations seen in giant flares  \citep{2005Natur.434.1107P,2005Natur.434.1098H}. Also, the  fact that radio peaks lead the X-rays  \citep{2020ApJ...898L..29M} is consistent with the prediction that acceleration occurs at the initial stage of the reconnection event, while the surrounding plasma is relatively clean of the pairs \citep{2019arXiv190103260L}.

\cite{2016MNRAS.462..941L,2019arXiv190103260L} discussed the overall constraint that observations impose on the FRB {\it loci} (this includes the required plasma density,  \Bf\ and bulk motion). They concluded that magnetars  generally satisfy those constraint.  

The \Bf\ plays the most important role in the generation and  propagation  regions. At the core of it is the laser non-linearity parameter  \citep{1975OISNP...1.....A}
\be
a \equiv \frac{e E_w}{m_e c \om}
\label{Akhiezer} 
\ee
where $E_w$ is the \Ef\ in the coherent wave, and $\om$ is the frequency (parameter $a$ is Lorentz invariant). In FRBs its value can be as large as $\sim 10^6$ \citep{2014ApJ...785L..26L,2016MNRAS.462..941L}.  In unmagnetized plasma the nonlinearity  parameter (\ref{Akhiezer}) is a dimensionless momentum of transverse motion of a particle in the EM wave. 
As a result, particles moving with \Lf\ $\gamma =a$ in the \EM\ experience ``normal' losses (this is a radiation reaction effect and is usually not taken into account). 

Three principal factors contribute to the suppression of the ``normal'' loses in  magnetar \mss: (i)  high \Bf, (ii)  relativistic outward motion of plasma; (iii) destructive interference of  waves  scattered by plasma particles.
In the  highly guide-field dominated regime, when cyclotron frequency is much larger than the wave frequency, the   parameter (\ref{Akhiezer})  loses its importance. The high \Bf\ in the inner regions of the \ms\ modifies particle motion in the coherent wave: instead of experiencing acceleration on the time scale of  quarter of a period $\sim 1 /\om$, the \Bf\ bends particle trajectory on time scale $\sim 1/\om_B \ll   1/\om$ \citep{2016MNRAS.462..941L,2020PhRvE.102a3211L,2021ApJ...918L..11L}. 

As
\cite{2019arXiv190103260L} argued, in the absence of strong guide-field a coherently emitting particle will lose energy on time scales shorter than the coherent low frequency wave. Thus large \Bf\ are {\it required }  in the emission region.
Particles that produce coherent EM waves in the inner parts of the \ms\ do not suffer ``normal'' (non-coherent) losses due to the high \Bf. 
But these losses may become important in the outer regions of the \ms. 

\cite{2021arXiv210807881B} argued that losses in the outer parts of the \ms\ will prevent escape of radio waves.  \cite{2021arXiv210807881B} argued that the combined effects of  enhanced scattering cross-section, and  what we call the ``normal losses'' would lead to strong energy dissipation of the wave.  In this contribution we show that, first, 
in the outer regions of the magnetars' \mss\ FRBs induce, via ponderomotive force, large \Lf\ motion  parallel to the direction of wave propagation. As a result,  the corresponding self-losses are negligible. Losses in the external fields (\eg\ the background \Bf\ and/or \Alfven wave propagating through the \ms) may occasionally be important and may lead to observable effects. 
 
 \section{Overall energetics}

For fiducial estimates,  consider an FRB  pulse of duration  $\tau = 1$ msec, coming from $D$= Gpc and producing flux $F_\nu$ = 1 Jy  at frequency of  $\nu=10^9$ Hz. The isotropic equivalent energy is then
\be
E_{iso} =  4 \pi D^2 \nu F_\nu \tau = 10^{39} {\rm erg}
\label{Eiso}
\ee
Let's normalize the  surface \Bf\ to the quantum field
\ba && 
B_{NS} = b_q  B_Q 
\nn &&
B_Q =  \frac{c^3 m_e^2}{e \hbar }
\label{BQ}
\ea 
Then the  energy needed to power the radio burst is contained within 
\be
R_f  = \left(  24 \pi \frac{ D^2 \nu F_\nu \tau}{B_{NS}^2}  \right)^{1/3}  = 
 1.5 \times 10^4 \times b_q^{-2/3}  {\rm cm} ,
\ee
about a football field.

Alternatively, assume that the total duration  of the burst is due to  "lateral" extension  of an 
active region located near the \NS\ surface, as its direction  of emission swings by an observer:
\be
R_f = \frac{\tau}{P}  {R_{NS}} 
\label{Rf}
\ee
where $P$ is the rotation period.  Emission is then confined into solid angle $\pi  \theta_{\rm b}^2$, $ \theta_{\rm b}=  (R_f/R_{NS})$. 
The real energetics is  $\pi (\tau/P)^2$ smaller,
\be
E_{real} =   4 \pi ^2 \frac{  D^2 \nu F_\nu \tau^3}{P}  = 3\times 10^{33 } P^{-2} {\rm erg}
\ee
for period measured in seconds. 
Such amount of energy  is contained within the layer of thickness
\be
\frac{(\Delta R)}{R_{NS}} =\frac{  32 \pi^2}{b_q^2}  \frac{ D^2 \nu F_\nu \tau}{ B_Q^2 R_{NS}^3}  =1 .5 \times  10^{-5} b_q^{-2}
\label{DeltaR}
\ee
about 15 centimeters only.

Radio emission is a tiny fraction of the total  energy budget - mostly the energy is spent on the accompanying X-ray emission \citep{2021NatAs...5..372R,2020Natur.587...59B,2020ApJ...898L..29M},  and presumably ejection of a plasma from the \ms.   Relation (\ref{DeltaR}) demonstrates that X-ray power  exceeding the radio by $10^5$ can be accommodated within the magnetospheric  model.

\section{Escape of FBRs from magnetars' magnetospheres}

\subsection{Lorentz transformations}
Consider an emission region moving with Doppler factor $\delta$. Let's denote the quantities measured in the plasma frame with primes. 
The
Lorentz transformations of frequency, flux, brightness temperature,  radiation energy  density and duration read \citep{2013LNP...873.....G,2019arXiv190103260L}
\ba && 
\nu' = \nu/\delta
\nn && 
F_{\nu'}' = F_{\nu}/\delta^3 
\nn &&
T_b' =T_b/\delta ^3
\nn &&
u_{rad}' = u_{rad} /\delta^4
\nn &&
\tau ' = \tau \delta
\ea

In the generation region the energy density of plasma particles  $u_p'$ should exceed the energy density of radiation $u_{rad}' $ (particles cannot emit more energy than they have). In addition, in magnetar \mss\ we expect that  the energy density is dominated by the  energy density of the    \Bf\ $ u_B' $. 
Thus,
\be 
u_B' \geq  u_p' \geq u_{rad}' 
\ee

Let us consider two case, first, when the emission region is scaled with the (Lorentz-boosted)  total duration of the burst, \S \ref{tauc}  and, second,  when duration is determined the lateral  extension of an active region  near the surface of a rotating  \NS\, \S \ref{taul}.

\subsection { Size of emission region  scaled $\propto \tau c$} 
\label{tauc} 
Given the observed flux $F_\nu$, duration $\tau$ and distance to the source $D$,
radiation energy densities evaluate to
\ba &&
u_{rad} = \frac{D^2 \nu F_\nu}{c^3 \tau^2}
\nn &&
u_{rad}' = \frac{D^2 \nu F_\nu}{c^3 \tau^2 \delta^4}
\ea
($u_{rad}' $ is expressed in terms  of the observed quantities).

Requirement $u_B' \geq u_{rad}' $ gives the estimate of the \Bf\ (a limit from below): 
\be
B_{eq}  \geq 2 \sqrt{\pi}  \frac{ \sqrt{ F_\nu \nu} D }{\tau c^{3/2} } \frac{1}{\delta^2} =2 \times 10^8  \times \delta^{-2}    {\rm G}
\ee
For  surface \Bf\  parametrized as (\ref{BQ})   this is satisfied for 
\be
\frac{r_{em}}{R_{NS}} \leq \frac{ b_q ^{1/3}  B_Q^{1/3} \sqrt{c} \tau ^{1/3} }{2^{1/3} \pi^{1/6} (F_\nu \nu)^{1/6} D^{1/3} }  \delta^{2/3} =60  \times  b_q ^{1/3}  \delta^{2/3} 
\label{rem}
\ee
Thus, emission must  be produced in the inner parts of the \ms.

If FRBs are produced during magnetic reconnection events, one might expect that in the dissipation region $u_B \sim  u_p$. 
If  the energy density in particle is smaller than in \Bf, $u_B = \sigma  u_p$, $\sigma \geq 1$, all the relation below will be modified as 
$b_q \rightarrow  \sigma^{-1/2} b_q$.   For example, condition (\ref{rem}) becomes
\be
\frac{r_{em}}{R_{NS}} \leq \frac{ b_q ^{1/3}  B_Q^{1/3} \sqrt{c} \tau ^{1/3} }{2^{1/3} \pi^{1/6} \sigma ^{1/6} (F_\nu \nu)^{1/6} D^{1/3} } =60
\times b_q ^{1/3}  \delta^{2/3}   \sigma ^{-1/6} 
\label{rem2}
\ee
In what follows we omit the factor $\sigma$, with understanding that $b_q  \equiv  \sigma^{-1/2} ( B_{NS}/B_Q)$.

The  \Ef\ in the wave and the  nonlinearity  parameter at the source evaluate to 
\ba  &&
E' = \frac{2 \sqrt{\pi}  D  (\nu F_\nu)^{1/2}}{ c^{3/2} \tau \delta^2}
\nn &&
a' = \frac{e E'}{2\pi m_e c \nu'} = \frac{e E}{2\pi m_e c \nu}=   \frac{ e   F_\nu^{1/2}}{ c^{5/2} \tau m_e \nu^{1/2}  \delta}
\ea
(The nonlinearity parameter $a$ is Lorentz invariant for plane  waves,  but not if expressed in terms of fluxes - the extra $\delta$-factor comes from aberration).

 Conventionally, {\it in the absence of strong external \Bf}, parameter $a$ is a dimensionless  transverse momentum of a particle in the \EM\ wave, $a \sim p_\perp/(m_e c)$.
If a particle oscillates in the \EM\ fields of the coherent wave with amplitude $a$,
then in addition to the energy losses to the emission of coherent waves it will also suffer Inverse Compton  (IC) and  synchrotron losses (if \Bf\ is present).   Those ``normal'' losses may dominated over the ``coherent'' losses. 

The normal loss time scale  can be estimated as  $\tau_{loss}$:
\ba &&
\frac{\gamma' m_e c^2}{ \tau_{loss}'} = \frac{e^4 u_{rad}' \gamma^{\prime, 2}}{ m_e^2 c^3}
\nn && 
\tau_{loss} = \delta \tau_{loss}'
\label{Losss}
\ea
(We use energy loss estimate in turbulent radiation field parametrized by energy density $ u_{rad}' $, not in a single coherent wave, Eq. (\ref{losscohe}). This is more appropriate in the wave production region)

We find
\be
\frac{\tau_{loss} }{\tau} =  \sqrt{\pi} \frac{ c^{21/2} m_e^4 \tau^2 }{ e^6 F_\nu^{3/2} \nu^{1/2} D^3}\delta^6 
= 10^{-10} \delta ^6
\ee
Thus,  in the absence of guiding \Bf\ normal losses will drain particles' energy before it has time to emit a coherent wave.

\cite[][formulated the corresponding condition in terms of the effective  brightness temperature $T_b$. The synchrotron/IC radiation decay times become shorter than pulse duration  for
\ba &&
\overline{T}_b \geq \frac{1}{k_B} \frac{ m_e ^{8/3} c^7}{ e^{10/3} \nu^{7/3} \tau^{2/3}} = 
10^{29}   \, {\rm K}
\nn &&
   \overline{T}_b \equiv    \left(  \frac{\Delta \nu}{\nu}  \theta_{\rm b}^2  \right)   T_b.
\label{Tmax}
\ea
$   \overline{T}_b$ incorporates  finite spectral bandwidth and anisotropy of the emission; $T_b$ is the observed  brightness temperature. 
]{2019arXiv190103260L} 

There is an important caveat to the above statements, which resolves the problem of large normal losses in the FRB production region \citep{2016MNRAS.462..941L,2019arXiv190103260L}. 
 In  the limit of large  guiding \Bfs, with $\om_B \geq 2\pi \nu'$ where $\om_B = e B_0/(m_e c)$ and $\nu'$ is the wave frequency in the source frame,  the nature of the particle's motion in the field  of the \EM\ wave changes: instead of   oscillations   in the \Ef\ of the wave with the dimensionless momentum $a$, a particle experiences slow drift with velocity $v_d/c \sim E'/B_0$ (here $E'$ is the \Ef\ of the wave, while $B_0$ is the external \Bf). This condition translates to
\be
B_0 \geq \frac{ m_e c^2}{e \lambda \delta} \approx 100  \delta^{-1}\, {\rm G} 
\label{Bc}
\ee
Thus, the  presence of high \Bfs\ in high brightness relativistic sources is required to avoid large radiative losses of coherently emitting particles to IC  and synchrotron processes.

\subsection {  Size of emission region $\propto ( \tau  /P) R_{NS}$}
\label{taul} 
Repeating  estimates of the \S \ref{tauc}, we find 
\ba && 
u_{rad} = \frac{D^2 \nu F_\nu}{ c  R_{NS} ^2}
\nn &&
B_{eq} =  2 \sqrt{2 \pi}  \frac{ (\nu F_\nu)^{1/2}  D }{ \sqrt{c} R_{NS}} =  8.6 \times 10^9 {\rm G}
\nn &&
\frac{r_{em}}{R_{NS}} \leq \frac{ b_q ^{1/3}  B_Q^{1/3} c^{1/6}  R_{NS}  ^{1/3} }{2^{1/2} \pi^{1/6} (F_\nu \nu)^{1/6} D^{1/3} } =17 b_q ^{1/3}
\ea
A very much similar result.

\subsection{Loses in the escape  region}

\subsubsection{Intrinsic losses in the field of  FRB pulse} 

As  an FRB pulse  propagates away from the star, the guiding field decreases.  At distances larger than $r_{em}$ the \EM\ fields in the wave are larger than the guiding field. This does not lead to dissipative   effects, since  the particle motion in the wave  still occurs on fast time scale given by the guide field - particles do not have time to acquire large energy from the  \Ef\ of the wave. 

For  a given surface field  the  condition $\om\leq \om_B$  (\ref{Bc}) implies radii
\be
\frac{r_r}{R_{NS}} \leq b_q ^{1/3} \left(  \frac{ e B_Q}{m_e c \om} \right)^{1/3} = 5 \times 10^3  b_q ^{1/3}
\label{rr} 
\ee

Condition $r_r = R_{LC}$  (\LC\ radius) is achieved for periods of 
\be 
P_r = \left ( \frac{4 \pi^2 b_q B_Q  }{m_e c^4 \nu} \right)^{1/3} \approx 1 {\rm s}
\ee
For \NSs\ with shorter period, $P  \leq P_r$, the frequency of the waves is    $\om \leq \om_B$ everywhere in the \ms: EM wave-particle interaction  is then suppressed everywhere  within the \ms. 

Let us next consider slower \NSs, $P\geq P_r$. In the outer regions of the \ms\ $ \om \leq \om_B$:  this is the conventional  laser-plasma interaction regime where guide field is not important. 

As  the waves propagate from their origin near the \NS, their amplitude decreases as $1/r$. 
At the radius $r_r$ we have
\be
a_r  = \frac{e D \sqrt{F_\nu} }{\sqrt{\pi} m_e c^{3/2} r_r \sqrt{\nu}}=  3.5 \times 10^3 b_q^{-1/3}
\label{ar} 
\ee

Let us first consider {\it single} particles losses  in an \EM\ with amplitude (\ref{ar}) (collective effects are discussed in \S \ref{LPM}). 
Importantly, as the FRB pulse propagates through the \ms\  particles experience ponderomotive force that accelerates them along the direction of the wave propagation to   $\gamma_\parallel = \sqrt{1+a^2} \approx a$ \citep[][]{1996PhPl....3.2183S}.  (A simple way to see this is to consider FRB pulse as a packet of highly relativistic, {\it nearly} luminal  \Alfven\  waves. Total energy is then conserved, transverse \Lf\ is of the order of the parallel one). 
This is a  {\it non-dissipative} process: the particle will give all  the parallel energy back to the pulse   on the descending part of the intensity envelope.

Acceleration to $\gamma_\parallel  = a$  assumes  pulse propagation parallel to the external \Bf. For  propagation perpendicular to the external field both the  particles and the fields will be accelerated \citep{2021arXiv210807881B}, but since ponderomotive force acts on time scales much longer that the cyclotron gyration in the dipolar field, no cyclotron motion will be excited (adiabatic  process - no Landau transitions):   no dissipation, energy exchange between FRB pulse and plasma is reversible.  Also, the energy of the dipolar \Bf\ within a volume  of a pulse ar $r_r$ is much smaller than the energy of the pulse, 
$4\pi r_r^2 (c \tau) B^2[r_r]/(8\pi) / E_{iso} = 4 \times 10^{-8} b_q^{2/3}$ - plasma will be accelerated to  $\gamma_\parallel  \gg 1$ in this case as well.

In the frame of the particle then the losses are controlled by 
\ba &&
\frac{\gamma'  m_e c^2}{ \tau_{loss}'} \approx  \frac{e^2}{c}  \gamma^{\prime, 4} \om^{\prime, 2}
\label{losscohe1}
\\ && 
\gamma'  \sim a
\nn && 
 \tau_{loss}' = \frac{m_e c^3}{a^3 e^2   \om^{\prime, 2}}
 \label{losscohe}
 \ea
 \citep[the coefficient on the rhs of (\ref{losscohe1})  is $2/3$ for circularly polarized wave and $1/8$ for linearly polarized][]{LLII}.
 
In the lab frame
\be
\tau_{loss} = \delta \tau_{loss}' = 8 \frac{ m_e c^3}{e^2 \om^2} = 2 \times 10^4 \, {\rm seconds}
\label{Losss1}
\ee
where we used $\delta =2 a$ - this is the effect of the ponderomotive acceleration.
Qualitatively, bulk ponderomotive acceleration ``kills'' all the loses approximately as  $\gamma_\parallel^{3}$:  smaller frequency ($ \propto  \gamma_\parallel^2$ in power),  and times scales stretched  ($ \propto \gamma_\parallel$). It is remarkable that the time scale (\ref{Losss1}) is independent of the pulse intensity parameter $a$.

 The loss time scale (\ref{Losss1}) is very long, much longer than the expected period  of a magnetar: FRB pulses do not suffer from strong radiative self-damping in the outer parts of the magnetar's  \mss.

\subsubsection{External losses} 

More dangerous are ``external'' losses:  particles moving  relativistically in the EM field of the FRB pulse with the total \Lf\ $\gamma \sim a_r^2 \sim 10^7$ (a product of parallel ponderomotively induced and perpendicular {\Lf}s)   may also loose energy due to radiative process  not  intrinsic to the the pulse, but external magnetospheric perturbations.  If the energy density of  external perturbations in the lab frame is $u_{\rm ex}$, the corresponding loss rate would be
\be
 \tau_{loss, ex}= \frac{m_e^3 c^5}{a_r^2 e^4 u_{\rm ex} }
 \ee
where the intensity parameter  $a_r$ is given by (\ref{ar}). 

Various estimates can be made to estimate   $ u_{\rm ex} $. For example, {\it if}  both the liner and the oscillating momenta of the particles induced by the FRB wave are mostly perpendicular to the underlying \Bf, the synchrotron loss time scale with $ \left. u_{\rm ex} = B^2/(8\pi) \right|_{r_r}$ is 
$\tau_{loss, ex} \sim 10^{-2} b_q^{2/3}$ seconds -  it may be important. (The corresponding photon energy is $\sim 500$ MeV.  This  extreme case  demonstrates that  external (to the proper FRB wave) effects  may lead to the dissipation of the FRB energy. But they do not have to. 

Another possibility is if a beam of  leptons  accelerated by the FRB pulse encounters \Alfven wave with the wavelength  $\lambda _A $ (somewhat) shorter that $r_r$, $\lambda _A = \eta_A r_r$,  $\eta_A  \ll 1$. For   $\eta_A  = 10^{-2} $,  this will produce a UV/soft X-ray  pulse with the same power as  the FRB proper, (\ref{Eiso}), a ``swan song'' of an FRB.
 The corresponding peak flux
 \be
 F = \frac{ \nu F_\nu}{\hbar \om_{A} } = 10^{-4} {\rm phot \, cm^{-2} s^{-1}}= 2 \times 10^{-16} {\rm erg \,  cm^{-2} s^{-1}}
\label{FX1}
 \ee
 $\om_A = 2\pi c/\lambda_A$, 
 could in principle be detected with sensitive instruments like Chandra' High Resolution Camera. The peak flux (\ref{FX1})  lasts only a millisecond \citep[for other estimates of FBRs'  signals see][]{2016ApJ...824L..18L}.

\section{Further suppression of loses due to LPM  effect.}
\label{LPM}

Finally let us comment on the  very applicability of a single particle approach in calculating radiative losses in a strong \EM\ wave.  
 In the  absence of the guide field the {\it single  particle} scattering cross-section in a strong wave with $a\gg 1$ is enhanced \citep[non-linear  Thomson scattering,][]{1975UsFiN.115..161Z,1993PhRvE..48.3003E},
\be
\sigma = \sigma_T (1 +a^2)
\ee
(or by $(1 +a^2/2)$ for linearly polarized wave).   \cite[Non-linear Thompson scattering  in the guide-field dominated regime has been considered by][]{2021arXiv210207010L}.

 Importantly, the criterium for    the single particle  interaction  (versus collective) is that the radiation formation length  is smaller than the distance between particles. Even though high energy particle emit short wave length $ \lambda_{em} = c/\om_{em}$ the radiation formation length is long $l_c \sim \gamma^2 \lambda_{em}$, Fig. \ref{Picture-of-RadiationFormation}. This surprising result (first discussed in applications to high energy particle scatterings)   is known as the  Landau-Pomeranchuk-Migdal (LPM)  effect \citep{1956PhRv..103.1811M}, historically  initially discussed by \cite{1961NucPh..24...43T}, also \cite{2021ApJ...918L..11L}.

In the case of motion in circularly polarized wave a particle rotates  with radius $c/\om'$, where $\om'$ is the frequency of the wave in the frame where the particle is at rest on average (guiding center frame). At each moment the  \Bf\ of the wave is counter aligned with the velocity while the \Ef\ provides the centrifugal force.  Single particle emitted frequency is $\sim a^3 \om'$, while the radiation formation length is $l_c' \sim ( c/\om' ) /a $. 
If inter-particle distance is smaller than $l_c '$ inference of waves emitted by different particles will  reduce the intensity of the scattered waves. In the continuous limit the emission is suppressed completely: a ring current does not emit.

In the frame of the ponderomotively accelerated plasma $\om' = \om/a$, hence  $l_c' \sim  c/\om $.
Since this is the length along particle motion, hence across wave propagation direction, it is the same in the lab frame, $l_c = l_c' =     c/\om $.
It will be typically   larger than the  inter-particle distance.

Thus, in the continuous limit the enhanced cross-section for the non-linear  Thomson scattering does not lead to any losses, no  extra scattering opacity. In this case the waves  scattered  from different particles  add constructively along the direction of wave propagation.  Just as in the conventional  cold plasma,  in the  continuous regime Thomson scattering by individual electrons   leads only to the modification of the dispersion relation.
 In the  non-linear regime, for circularly polarized light the photon dispersion  becomes 
\cite[][Eq. 8.1.4.1]{1975OISNP...1.....A}
\be
\om^2 = (k c)^2 + \om_p^2 /\sqrt{1+a^2} 
\label{omegofk} 
\ee


Much more  interesting, and possibly  more  dangerous for wave propagation, could be  ponderomotive effects  for  linearly polarized waves  \citep[][]{1986PhR...138....1S}. 

\section{Discussion}
In this Letter we consider escape of high brightness coherent FRB  emission from magnetars' \mss. We come to a different conclusion than \cite{2021arXiv210807881B}: FRB pulses  typically escape. 
Radiative losses of particles moving in the field of an FRB    could  in principle be important, but for a very restricted set of conditions. First, it requires a fairly  slow \NS\, with period $P \geq 1$ sec. Second, even for {\NS}s with longer periods the single particle emission is  suppressed by the non-dissipative ponderomotive acceleration of the background plasma by the  incoming FRB pulse. Third, nonlinear Thomson scattering in a strong EM wave  is   suppressed by the LPM effects (long radiation formation  length and the corresponding destructive interference):  FRB pulse propagates not through a collection of single scatterers but through a continuous medium:  wave propagate non-dissipatively, with only  a slightly modified dispersion. 
 
 In some case  radiation effects of FRB-accelerated particles on external perturbation may be important and may lead to observable 
 weak UV/soft X-ray pulse that could in principle be detected by the sensitive instruments like Chandra.

 This work had been supported by 
NASA grants 80NSSC17K0757 and 80NSSC20K0910,   NSF grants 1903332 and  1908590.
I would like to thank Andrei Beloborodov, Henry Freund  and Victoria Kaspi for discussions. 

\section{Data availability}
The data underlying this article will be shared on reasonable request to the corresponding author.

\bibliographystyle{apj}

  \bibliography{/Users/maxim/Home/Research/BibTex}

\end{document}